\newcommand{\fig}[3]{
 \begin{figure}[!hbt]
  \begin{center}
   \begin{minipage}{0.85\textwidth}
    \centering{\includegraphics[#1]{#2}}
   \end{minipage}
   \caption{\label{#2}\small{#3}}
  \end{center}
 \end{figure}
}
\newcommand{\tab}[3]{
 \begin{table}
  \begin{center}
   \caption{\label{#1}\small{#2}}
   \begin{minipage}{0.85\textwidth}
    \centering{#3}
   \end{minipage}
  \end{center}
 \end{table}
}
\documentclass{article}
\usepackage{amsfonts, amsmath, amssymb, amsthm, centernot, cite, enumitem}
\usepackage{algorithm}
\usepackage[noend]{algpseudocode}
\usepackage{diagbox}
\usepackage{graphicx}
\usepackage{multirow}
\usepackage{tikz, xcolor, xspace}
\makeatletter
\renewcommand{\fnum@figure}{FIG. \thefigure}
\renewcommand{\fnum@table}{TABLE \thetable}
\renewcommand{\thetable}{\Roman{table}}
\newcommand\et{\textit{et al.}\xspace}
\newcommand\su[1][]{$\mathsf{su}$#1\xspace}
\newcommand\circled[1]{\tikz[baseline=(char.base)]{%
 \node[shape=circle,draw,inner sep=0.5pt](char){#1};}}
\def\BState{\State\hskip-\ALG@thistlm}
\makeatother
\begin{document}
\title{Improved OpenCL-based Implementation of Social Field Pedestrian Model}
\author{
Bin Yu
\footnote{Corresponding author, Tongji University, P.R.C.,
\texttt{by@tongji.edu.cn}}
\and
Ke Zhu
\footnote{Tongji University, P.R.C.,
\texttt{1213235421@qq.com}}
\and
Kaiteng Wu
\footnote{Tongji University, P.R.C.,
\texttt{kaitengwu@163.com}}
\and
Michael Zhang
\footnote{University of California at Davis, U.S.A.,
\texttt{hmzhang@ucdavis.edu}}
}
\date{}
\maketitle
\begin{abstract}

Two aspects of improvements are proposed for the OpenCL-based
implementation of the social field pedestrian model.
In the aspect of algorithm, a method based on the idea of
divide-and-conquer is devised in order to overcome
the problem of global memory depletion
when fields are of a larger size.
This is of importance for the study of finer pedestrian
walking behavior, which usually requires larger fields.
In the aspect of computation, the OpenCL heterogeneous
framework is thoroughly studied.
Factors that may affect the numerical efficiency
are evaluated, with regarding to the social field model
previously proposed.
This includes usage of local memory, deliberate patch of
data structures for avoidance of bank conflicts, and so on.
Numerical experiments disclose that the numerical efficiency
is brought to an even higher level.
Compared to the CPU model and the previous
GPU model, the current GPU model can be at most
$71.56$ and $13.3$ times faster respectively
so that it is more qualified to be a core engine
for analysis of super-large scale crowd.

\noindent{\bf Keywords}:
Algorithm,
Heterogeneous Parallel Computing,
OpenCL,
Pedestrian Flow
\end{abstract}

\section{Literature Review}
\label{s: literature review}

Since NVIDIA first proposed the term ``Graphics Processing Unit'',
known as GPU, in the year of 1999, GPU has grown into
a heterogeneous parallel computing architecture,
which is often used to solve complicated
scientific and engineering problems.
The two most commonly used platforms are OpenCL and CUDA.
Scholars from different disciplines reported
successful applications of the two, like
\cite{cao.et.al, veronese.krohling} in mathematics,
\cite{bach.et.al} in physics,
\cite{pu.et.al, iwai.et.al, harish.narayanan,
zhu.et.al} in computer science,
\cite{garrett.et.al, khokhlov.et.al, molero.et.al,
komatitsch.et.al} in seismic engineering,
\cite{chang.et.al, ligowski.rudnicki}
in bioinformatics,
\cite{wang.et.al} in communication,
\cite{callico.et.al,
keck.et.al, pan.et.al, scherl.et.al} in image processing,
and so on.

A crowd is fundamentally a many-body system that would
require quite a lot of computation efforts for analysis.
GPU should naturally provide a good solution to the
problem in terms of numerical computation.
Unfortunately, the research on GPU in the field of
crowd simulation is far behind other disciplines.
Before the authors, only a few scholars have been engaged
in relevant research%
\cite{was.et.al, mroz.was, dutta.et.al, rahman.et.al}.
Furthermore, as far as the final results are concerned,
their research findings are not exciting and
suggest a space for further improvement.
As referred by Molero \et\cite{molero.et.al},
developing a scalable and portable GPU parallel model
is a challenge.
Especially, in order to fully utilize the power of GPU,
owing a suitable architecture is the key.
In other words, for a math model, it can be well mapped
into GPU only when its logic architecture is appropriate.
With realizing the point, the authors decided to propose
a field-based pedestrian model first after consideration%
\cite{yu.et.al-1}.
In addition to the better modeling of crowd dynamics,
the proposed continuous model has the advantage of easy
discretization so that a discrete version and a later
OpenCL-based implementation were developed.
Yu \et\cite{yu.et.al-2} reported that this can bring
an at most $30.8$ times speedup with comparison to the CPU model.

The paper is the follow-up research and tries to improve
the work of \cite{yu.et.al-2} in two aspects.
Firstly, it develops a method based on the idea of
divide-and-conquer to solve the problem of
global memory depletion when fields have
a large geometric size.
This is key as now it is possible to analyze
super-large scale crowd's finer walking behavior.
Secondly, potential factors affecting OpenCL are
thoroughly considered in order to further improve
the numerical efficiency.
The left content is organized as the following.
A brief of the continuous and discrete social field
pedestrian models is presented at first.
The discussion of introduced improvements comes next.
Then conducted numerical experiments are exhibited.
The conclusion is given in the end.

\section{Continuous and Discrete Models}
\label{s: continuous and discrete models}

Firstly, in order to avoid unnecessary confusion with
the cellular automata, the term of space unit
abbreviated as \su is used to express the minimal discrete
space.  To save space, only a brief is given.
Interested readers can refer to
\cite{yu.et.al-1, yu.et.al-2} for a detailed
description of the models.

In the continuous model,
a pedestrian's physical movement is simulated as
a response to the pedestrian's subjective perception
of the objective environment.
The objective environment is represented by force incurred
by presumed fields.  To model various practical phenomena,
total five kinds of fields are introduced, among which
omnidirectional attractive and repulsive fields are
to model influence of static openings and obstacles.
Directional attractive and repulsive fields and
recurrent repulsive fields are all to model influence
due to neighboring pedestrians' movement.
But their evolution laws are different as
they are targeting at walking behavior observed under
different density regimes.
Another point worthy of mention is introduction of
the concept of regulation function.
Through regulation functions, the objective environment
around a pedestrian $p$ represented by force can be
adjusted correspondingly to form the dynamic subjective or
perceived environment that is used to determine $p$'s
next movement.  In this way, pedestrians' intelligence
can be well considered.  Additionally it is stressed
that the \textit{local} density instead of
the well-known macroscopic one should be used to reflect
pedestrians' biased perception of the environment.

Except allowing the continuous model to own a larger degree
of freedom, using the concept of field also leads to
a straightforward discretization.
As shown in the work of \cite{yu.et.al-2}, a discrete model
was derived.  Especially, the concept of walk period was
introduced.  Using the concept, the variance of walking velocities
of pedestrians can be well studied under the assumption that
pedestrians' maximal speed is 1 \su per tick.
It should be noted that, for the discrete model,
the assumption $v_{max} = 1$ is not insignificant, but
one of the key points.
More importantly, the method will not add additional
complexities to the underlying logic,
but it should be emphasized that the concept
is related with some form of space fineness.
Therefore, in the discrete model, pedestrians are allowed to
occupy more than one \su and the behavior of jostling can
be studied.
In the meantime, to ensure that, for a pedestrian $p$,
one and only one \su will be $p$'s center, it is ruled
that $p$'s width and height can be different but must be
odd \su{(s)} like $1$, $3$, $5$, ..., and so on.

\section{Algorithm Related Improvement}
\label{s: algorithm related improvement}

\subsection{Architecture of OpenCL-based Computation Model}
\label{s: architecture of opencl-based computation model}

\begin{algorithm}
\caption{OpenCL-based GPU Model}\label{a: opencl based gpu model}
\begin{algorithmic}[l]
\Procedure{Main}{}
\State $p \gets \text{simulation period}$
\State $t \gets 0$
\While{$t < p$}
\State \emph{k-1.} initialize the temporary storage;
\State \emph{k-2.} determine pedestrians' next movement;
\State \emph{k-3.} vote which pedestrian should occupy unoccupied \su{(s)};
\State \emph{k-4.} perform pedestrians' next movement;
\State \emph{k-5.} write cached changes back;
\State $t \gets t + 1$
\EndWhile
\EndProcedure
\end{algorithmic}
\end{algorithm}

Algorithm~\ref{a: opencl based gpu model} lists
the architecture of the OpenCL-based model previously developed.
As indicated, five sub-jobs will be repeated orderly
at every simulation tick.
In \emph{k-1}, for each \su, a work-item will be assigned
to initialize the temporary storage allocated in the global memory space.
In \emph{k-2}, for each pedestrian, a work-item will be assigned
to determine the pedestrian's next movement.
In \emph{k-3}, for each \su, a work-item will be assigned
to vote which pedestrian among the candidates should occupy
if the \su is unoccupied.
In \emph{k-4}, for each pedestrian, a work-item will be
assigned.  If the pedestrian's next movement is not still,
the assigned work-item then checks whether the pedestrian is
the one to occupy for the \su{(s)} to be occupied.
If yes, the pedestrian will be physically moved and
relevant changes will be buffered.
Finally, in \emph{k-5}, for each \su, a work-item will be assigned
to write the changes buffered in \emph{k-4} back to
the global storage, which is camouflaged as a 3-D image.

When the discrete model was mapped into the OpenCL heterogeneous
computing framework, mechanisms were introduced, among which
two used to avoid atomic functions like \texttt{atomic\_add} etc
are worthy of little words.
In real applications, using atomic functions is the direct way for
solving competition among work-items.
On the other hand, it should be noted that atomic functions would
significantly harm the computation performance, especially when
a global memory is being manipulated.
Thus, to achieve a better performance, people will struggle to
avoid atomic functions, if possible, even though this may generally
require an overhaul.
For our problem, competition could occur in the
following two situations.

\begin{itemize}
\item[\textbf{s-1}.] Fields stored in the shared 3-D image
are updated due to pedestrians' movement.

Competition is solved through the concept of strength fan-out.
It is observed that, for one memory storage place in the shared
3-D image, once a field's discrete geometry space is determined,
the set of \su{s} can be computed beforehand so that
the memory storage place's content will be affected only if
the field's central \su belongs to the computed set.
Furthermore, although the set of \su{s} computed would change
if the field is moving,
the number of \su{s} keeps unchanged so that the concept of
strength fan-out is introduced.
This can be illustrated by examining a recurrent repulsive field
that is locating at the origin and whose discrete geometry space
is $7\times 7$ \su{s}.
Figures~\ref{discrete-sect.eps} and \ref{discrete-strength.eps}
give the discrete field strength incurred.
According to figure~\ref{discrete-sect.eps}, it is found that
the presumed recurrent repulsive field's strength fan-out is 6
and the set of computed \su{s} contains $(-1, -1)$, $(-1, 0)$,
$(0, 0)$, $(1, 0)$, $(0, 1)$, and $(1, 1)$.
Figure~\ref{update-race.eps} demonstrates that,
if the field's central \su belongs to the computed set,
the sect index of incurred discrete strength at
the \su $(2, 2)$ is always 1, meaning that
the same global memory address will be accessed.

With the concept of strength fan-out,
a big enough global memory space can be allocated
beforehand to buffer field strength related
changes made in \emph{k-4},
which will be written back to
the shared 3-D image all at once in \emph{k-5}.
In this way, no atomic function is required.

\fig{width=0.45\textwidth}{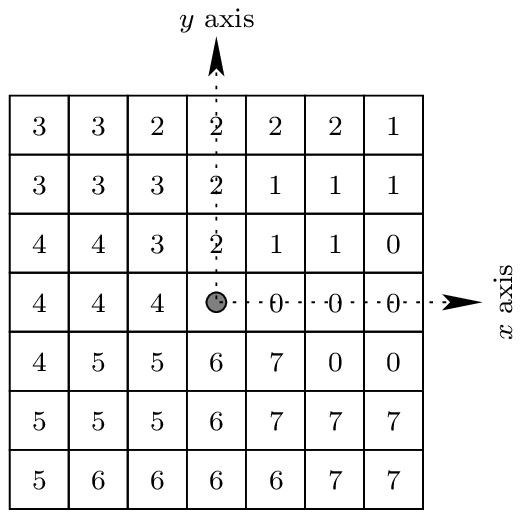}
{This exhibits sect indexes of field strength incurred by
the presumed recurrent repulsive field with
$k = 1$ and $\alpha = -0.5$.
A same sect index means that the same memory storage place
is to be changed.}

\fig{width=0.45\textwidth}{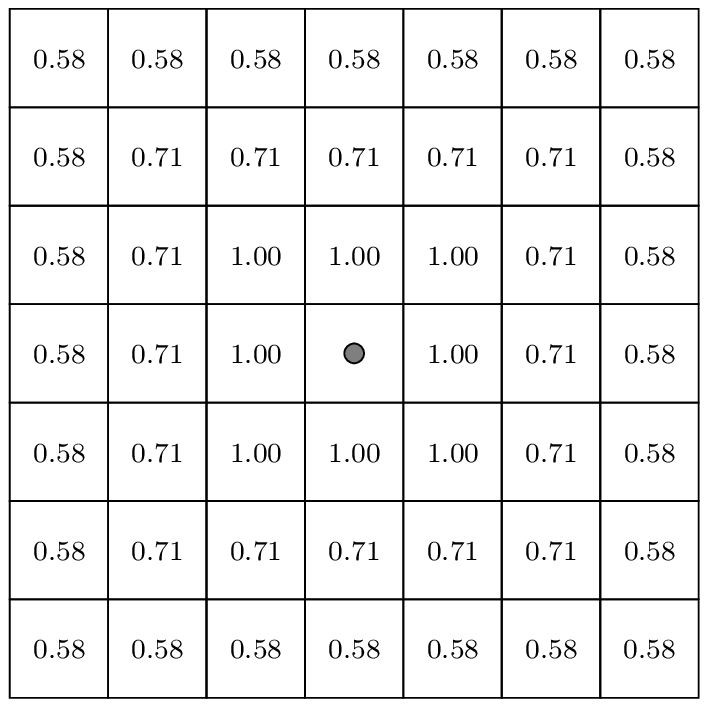}
{This exhibits scalar values of field strength incurred by
the presumed recurrent repulsive field.}

\fig{width=0.5\textwidth}{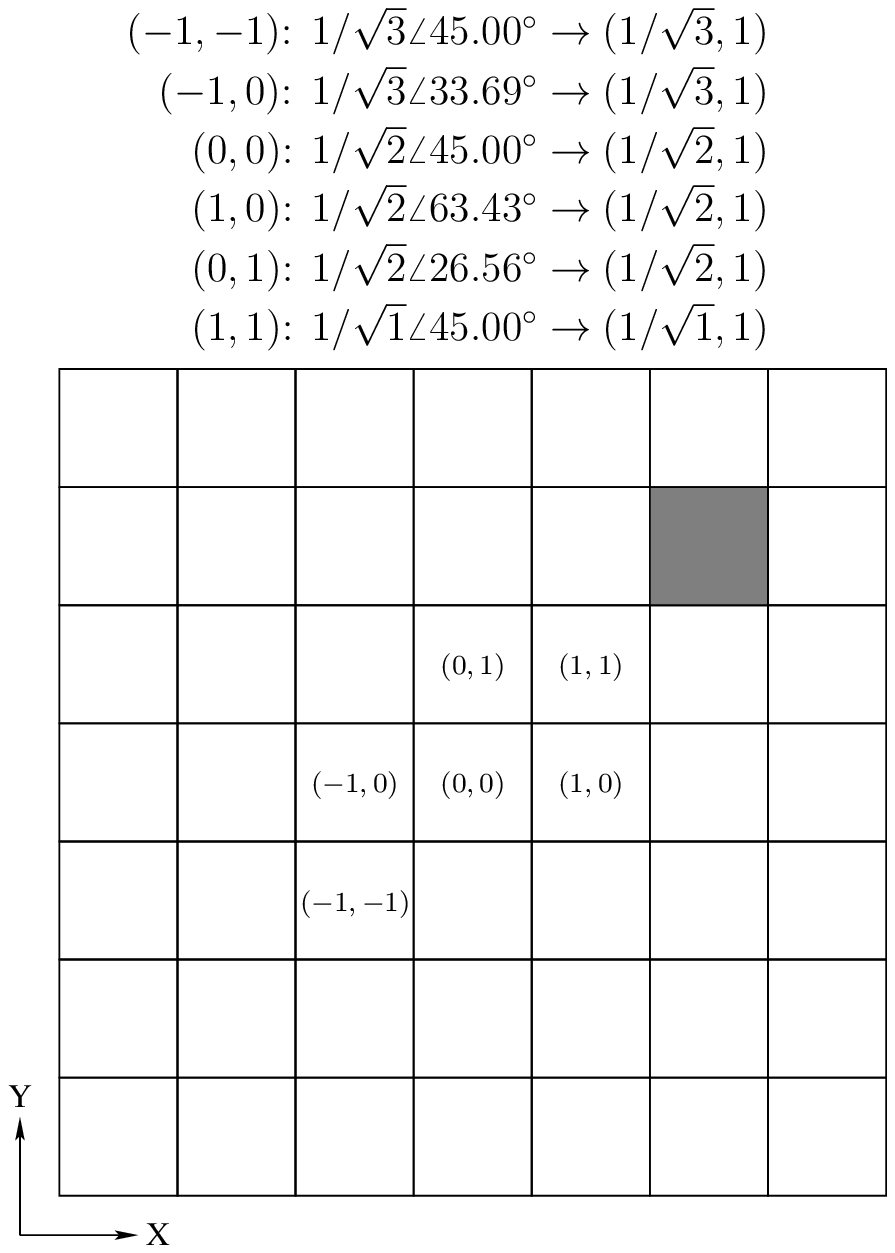}
{This exhibits, for the presumed recurrent repulsive field,
6 \su{s} exist so that, if the field's central \su is one of them,
the field can incur discrete strength at the \su $(2, 2)$
with the sect index value being $1$.}

\item[\textbf{s-2}.] Pedestrians compete with each other
for empty \su{(s)}.

Competition is solved by the observation that at most
$8$ pedestrians will participate the occupancy competition of
one \su under the assumption $v_{max} = 1$.
A register-vote mechanism is adopted.
An enrollment container that can hold at most $8$ pedestrians
will be allocated for each \su.
For an empty \su, pedestrians trying to occupy the \su should
register first at the \su's enrollment container (\emph{k-2}).
Later, an election among the registered pedestrians will be hold
to determine who should occupy the \su (\emph{k-3}).
\end{itemize}

\subsection{Technical Problems Suffered}
\label{s: technical problems sufferred}

The foregoing methods both follow the same idea of
preventing competition in expense of memory space,
but a big difference does exist.
For the method used to solve competition occurring
in \textbf{s-2}, an upper-bound limit with regarding
the required memory space exists since at most $8$
pedestrians will compete with each other
for occupancy of a \su.
Unfortunately, this is no more valid for the method
used to solve competition occurring in \textbf{s-1}.
When a field's discrete geometry space becomes larger,
the corresponding strength fan-out increases.
So is the required memory space.
Table~\ref{t: strength fan-out} lists strength fan-outs
when different geometries are assumed for
the presumed recurrent repulsive field.

\tab{t: strength fan-out}{List of strength fan-outs}{
\begin{tabular}{c|ccccccc}
\backslashbox{$\scriptstyle wd$}{$\scriptstyle ht$} & 1 & 3 & 5 & 7 & 9 & 11 & $\cdots$\\\hline
1 & 0 & 1 & 2 & 3 & 4 & 5 & \multirow{6}{*}{$\vdots$}\\
3 & 1 & 1 & 2 & 5 & 8 & 11\\
5 & 2 & 3 & 4 & 6 & 10 & 14\\
7 & 3 & 6 & 6 & 6 & 10 & 14\\
9 & 4 & 8 & 9 & 10 & 11 & 15\\
11 & 5 & 11 & 14 & 15 & 15 & 15\\
$\vdots$ & \multicolumn{6}{c}{$\cdots$} & $\ddots$\\
\end{tabular}}

In general, a field's discrete geometry will take a value
shown in table~\ref{t: strength fan-out}, thus is not large,
but situations where a large geometry is used do exist.
Firstly, a study of pedestrians' finer walking behaviors
generally requires larger fields so that distant effect
can be well considered.
Secondly, pedestrians are allowed to occupy more than one \su
so that fields should be scaled up correspondingly.
For the presumed recurrent repulsive field,
table~\ref{t: strength fan-out 2} lists
the strength fan-outs for different geometries.
As shown in figure~\ref{strength-fanout-ratio.eps},
the strength fan-out is increasing at a dramatic rate.

\tab{t: strength fan-out 2}
{This exhibits strength fan-outs when different geometries
are assumed for the presumed recurrent repulsive field.
For a ratio, the corresponding geometry's width and height
will both equal to $7 * ratio$.}%
{\begin{tabular}{c|cccccc}\hline
ratio & 1 & 3 & 5 & 7 & 9 & 11\\
\textit{strength fan-out} & 6 & 61 & 164 & 328 & 535 & 808\\
$m_{recur}$; byte & 192 & 1952 & 5248 & 10496 & 17120 & 25856\\
$M$; GB & 0.7 & 7.3 & 19.6 & 39.1 & 63.8 & 96.3\\\hline
\end{tabular}}

\fig{width=0.6\textwidth}{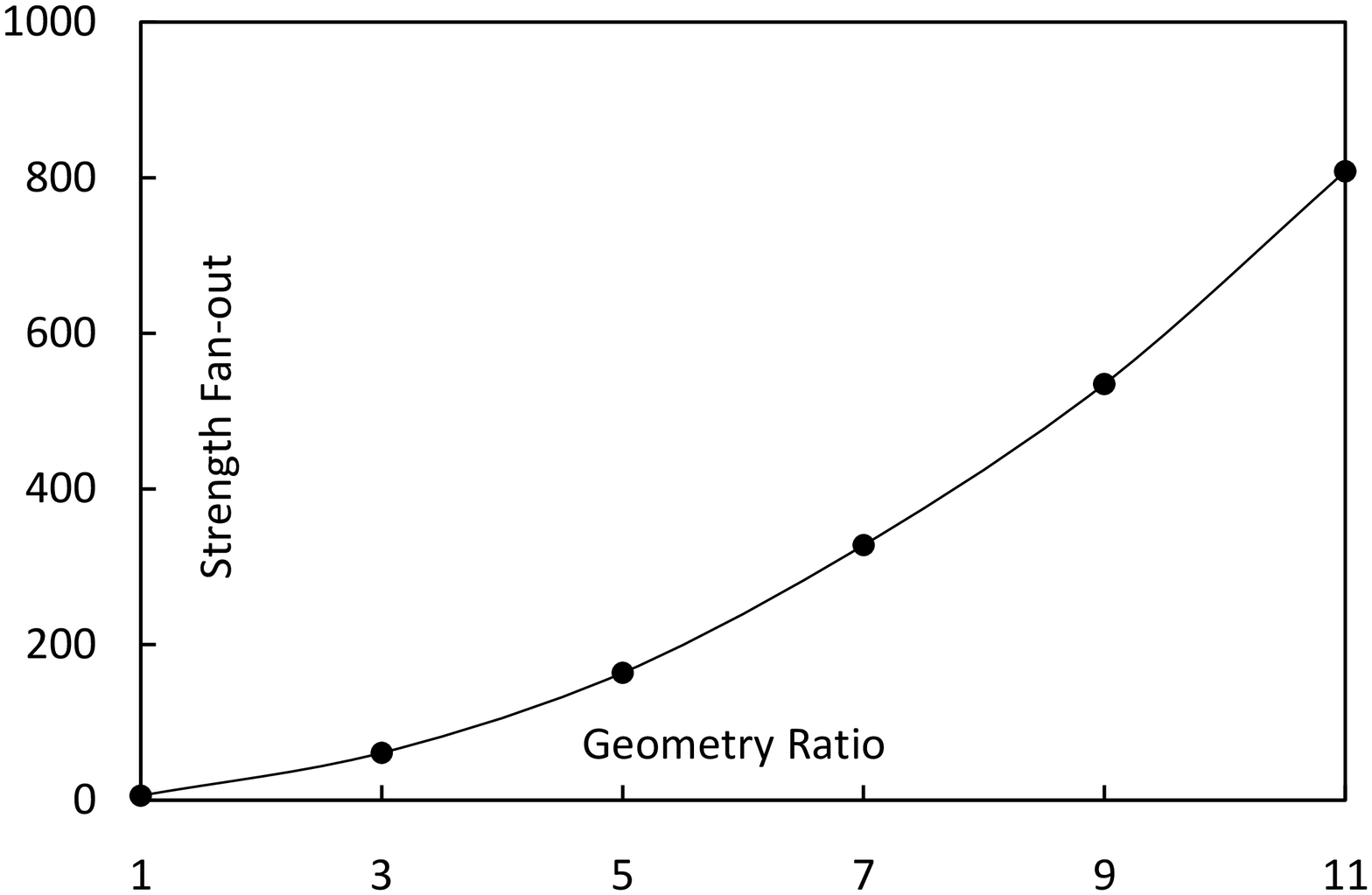}
{Geometry Ratio vs. Strength Fan-out}

The OpenCL-based implementation was developed for simulation of
super-large scale crowd.  Let us examine the memory space
required to solve competition occurring in \textbf{s-1}
if a half million population is simulated.
To keep the discussion simple,
the following assumptions are made.
The whole discrete space is $1000\times 1000$
and pedestrians are all of a geometry $1\times 1$ \su.
Thus a half million population would actually mean that
the macroscopic density is $0.5$.
All fields' geometry is same, so is their strength fan-out
denoted as \textit{SF}.
For each \su, the memory space $m_{recur}$ required to store
recurrent repulsive fields is then $\text{\textit{SF}} * 8 * 4$
and the total memory space needed is
$m = m_{attra} + m_{repul} + 2 * m_{recur}$.
Lastly, multiplying $m$ by the number of \su{s} gives
the total memory space $M$ needed.
Table~\ref{t: strength fan-out 2} also lists values of
$m_{recur}$ and $M$.
As shown, when fields' geometry is $21\times 21$,
$M$ already reaches to about $7.3$ GB,
not to say even larger sizes.

\subsection{Solution Illustration}
\label{s: solution illustration}

The former methodology is to cache all of the changes at a time
and algorithm~\ref{a: one-step sum} lists the pseudo-code
for recurrent repulsive fields.
For directional attractive and repulsive
fields, the algorithms are almost same.
However this may cause the foregoing memory depletion problem.

\begin{algorithm}
\caption{One-Step Sum For Recurrent Repulsive Fields}\label{a: one-step sum}
\begin{algorithmic}[1]
\Procedure{One-Step Sum in \emph{k-4}}{float cache[]}
 \ForAll{$c \in \text{\su{s}}$}
  \State $\text{\texttt{float *$s$}} \gets \text{\Call{FindCache}{cache, $c$}}$
  \State \Call{ComputeAndCacheChanges}{$s$}
 \EndFor
\EndProcedure
\item[]
\Procedure{One-Time Sum in \emph{k-5}}{float cache[]}
 \ForAll{$c \in \text{\su{s}}$}
  \State $\text{\texttt{float *$s$}} \gets \text{\Call{FindCache}{cache, $c$}}$
  \State $sum \gets 0$
  \For{$i = 0$ to $8 * \text{\textit{strength fan-out} - 1}$}
   \State $sum \gets sum + s[i]$
  \EndFor
  \State \Call{WriteBack}{$c$, $sum$}
 \EndFor
\EndProcedure
\end{algorithmic}
\end{algorithm}

\begin{equation}\label{eq: one vs. multi summation}
\sum_{i=0}^{n-1}a[i] \overset{\scriptscriptstyle n=K*m}{=} \sum_{j=0}^{K-1}\sum_{i=0}^{m-1}a[j*K+i]
\end{equation}

The proposed solution is based on the observation that
what is really cared is the final change summed up.
The summation way is of less importance.
As the aforementioned one-step summation may cause
the problem of memory depletion, the multi-step summation
can be used instead.
This is mathematically equivalent to Eq.~\ref{eq: one vs. multi summation}.
When $n$ is too large, it is fine to divide it into the multiplication
of two numbers, i.e. $n = K * m$.  Especially, with keeping one number
constant, an upper bound of memory consumption can be set up.

\begin{algorithm}
\caption{Multi-Step Sum For Recurrent Repulsive Fields}\label{a: multi-step sum}
\begin{algorithmic}[1]
\Procedure{Multi-Step Sum in \emph{k-4}}{float cache[]}
 \ForAll{$c \in \text{\su{s}}$}
  \State $\text{\texttt{float $s$[$K$]}} \gets \text{\Call{FindCache}{cache, $c$}}$
  \State $m \gets 8 * \text{\textit{strength fan-out}} / K$
  \For{$i = 0$ to $m - 1$}
   \State \Call{ComputeAndCacheStepChanges}{$s$, $i$}
  \EndFor
 \EndFor
\EndProcedure
\item[]
\Procedure{Multi-Step Sum in \emph{k-5}}{float cache[]}
 \ForAll{$c \in \text{\su{s}}$}
  \State $\text{\texttt{float $s$[$K$]}} \gets \text{\Call{FindCache}{cache, $c$}}$
  \State $sum \gets 0$
  \For{$i = 0$ to $K-1$}
   \State $sum \gets sum + s[i]$
  \EndFor
  \State \Call{WriteBack}{$c$, $sum$}
 \EndFor
\EndProcedure
\end{algorithmic}
\end{algorithm}

The proposed methodology's pseudo-code is given in
algorithm~\ref{a: multi-step sum}.
With comparison to the one shown in algorithm~\ref{a: one-step sum},
one additional loop between lines 5 and 6 is first noticed,
which actually performs the inner summation of
Eq.~\ref{eq: one vs. multi summation}.
More importantly, through the additional loop,
the loop count of the one between lines 11 and 12 is
bound to $K$.
Correspondingly, this would limit the memory space required,
i.e. \texttt{float[$K$]} in lines 3 and 9.
In the up-to-date implementation, $K$ is allowed to be
$2$, $4$, $8$, and $16$ in order to use
the built-in geometric function - $dot$ product.
Sometimes $K$ may not divide $n$, i.e. $K \centernot| n$.
To solve, $n$ can be extended to a multiplier of $K$ by
appending corresponding zero strength.

\section{Computation Related Improvement}
\label{s: computation related improvement}

\subsection{Structure of OpenCL}
\label{s: structure of opencl}

\fig{width=0.75\textwidth}{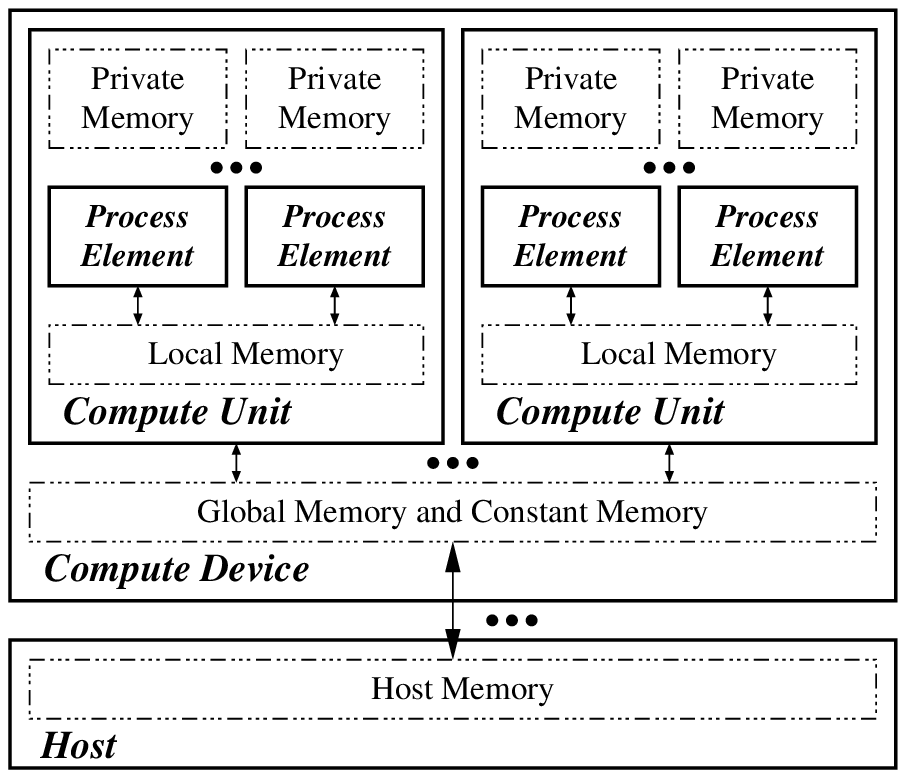}{OpenCL Architecture}

Computation techniques are developed with the OpenCL architecture
being kept in mind.  Therefore it is worthy to briefly illustrate
the architecture at first, as shown in figure~\ref{architecture.eps}.
At the hardware level, an OpenCL computing environment is composed of
a host and one or more compute devices.
A compute device is composed of multiple compute units,
each of which is further composed of multiple process elements.
In the level of execution, compute unit and process element
are also named work-group and work-item respectively.
To work with OpenCL, the general work flow chart is as follows.
Firstly, computation tasks to be fulfilled are
programmed as kernels, using the OpenCL C dialect.
Secondly, tasks are submitted to the command queue.
The OpenCL computing environment then takes
the full control of submitted tasks.
For example, at a suitable time, it will pick an apt
task and execute with appropriate global and local work sizes.
In order to improve the numerical efficiency,
OpenCL provides both task and data level of parallelism.
The task level of parallelism is embodied by the fact
that more than one task can be executed concurrently
as multiple compute units exist.
OpenCL provides the SIMD processing fashion for
the data level of parallelism, which can be viewed in
two sub-levels.  Logically, when a task is executed,
the kernel representing the task will be run literally
at the same time by work-items distributed among
more than one work-group.
Therefore a work-group is a bunching of work-items,
but not the most basic one, which is instead \textit{warp}
(Note: AMD calls this \textit{wavefront}.)
A warp is the smallest execution unit of code so that
the same machine instruction will be emitted to
all of work-items in it.
In this sense, the mechanism of warp provides the physical
level of data parallelism.

In addition, the OpenCL memory model is noteworthy too,
which is also exhibited in figure~\ref{architecture.eps}.
Firstly, data stored in the host memory can not be
directly accessed by compute devices and must be transferred
to the global memory beforehand through relevant OpenCL functions.
The global memory and constant memory can be accessed by
all process elements in the same compute device.
In addition, each compute unit has its own local memory
that is accessible to all process elements in it.
Lastly, each process element has the private memory,
which can be accessed only by the process element itself.
Except access privileges, the memory spaces are distinguished
in aspects of band width, capacity, and so on.
In terms of capacity, the global memory is largest and
reaches to the level of gibibyte.
Next is the local memory, which is generally in the level of kilobyte.
The private memory is least that is generally 1 kilobyte or less.
On the other hand, in terms of band width, the private memory is largest,
then is the local memory.  And the global memory is least.

\subsection{Computational Techniques}
\label{s: computational techniques}

With OpenCL's architecture being thoroughly investigated,
the following aspects are considered to improve
the numerical efficiency.

\begin{description}[leftmargin=0pt,font=\normalfont]

\item[\underline{\textit{Data Bandwidth}}]
In the former implementation, all computation work
is fulfilled by directly manipulating data
in the global memory after the transferring
from the host memory.
However this way does not fully consider
the OpenCL memory model's structure.
In OpenCL devices, the local memory is on-chip
and close to the process elements,
thus is much faster than the global memory.
By saying that, a more appropriate way to work with
the memory model is as follows.
Firstly, data is transferred from the host memory
to the global memory.
Secondly, for each work-group, the portion of data
to be worked by the work-group is copied to
the local memory through asynchronous copying functions
like \texttt{async\_work\_group\_copy} etc.
Once computation work is done,
data may be transferred back to synchronize
the one stored in the host memory, if necessary.

\item[\underline{\textit{Concurrency}}]
Formerly it is programmed so that tasks are almost executed
in the order of submission.
In addition, there is a strict execution sequence
between CPU and GPU so that CPU has to wait GPU to
finish all of the submitted tasks before starting to
process routine work.
Now the whole framework is re-organized so that
1. many times, more than one task can be executed simultaneously;
2. the strict sequence between CPU and GPU is weakened
to a large extent and even disappears in some cases.
Figure~\ref{workflow.eps} exhibits the flow chart of jobs
submitted.  Further, to see in which time frame each submitted job
is executed, a time profiling is fulfilled to one of numerical
experiments conducted, as shown in figure~\ref{percent.eps}.

\fig{width=0.8\textwidth}{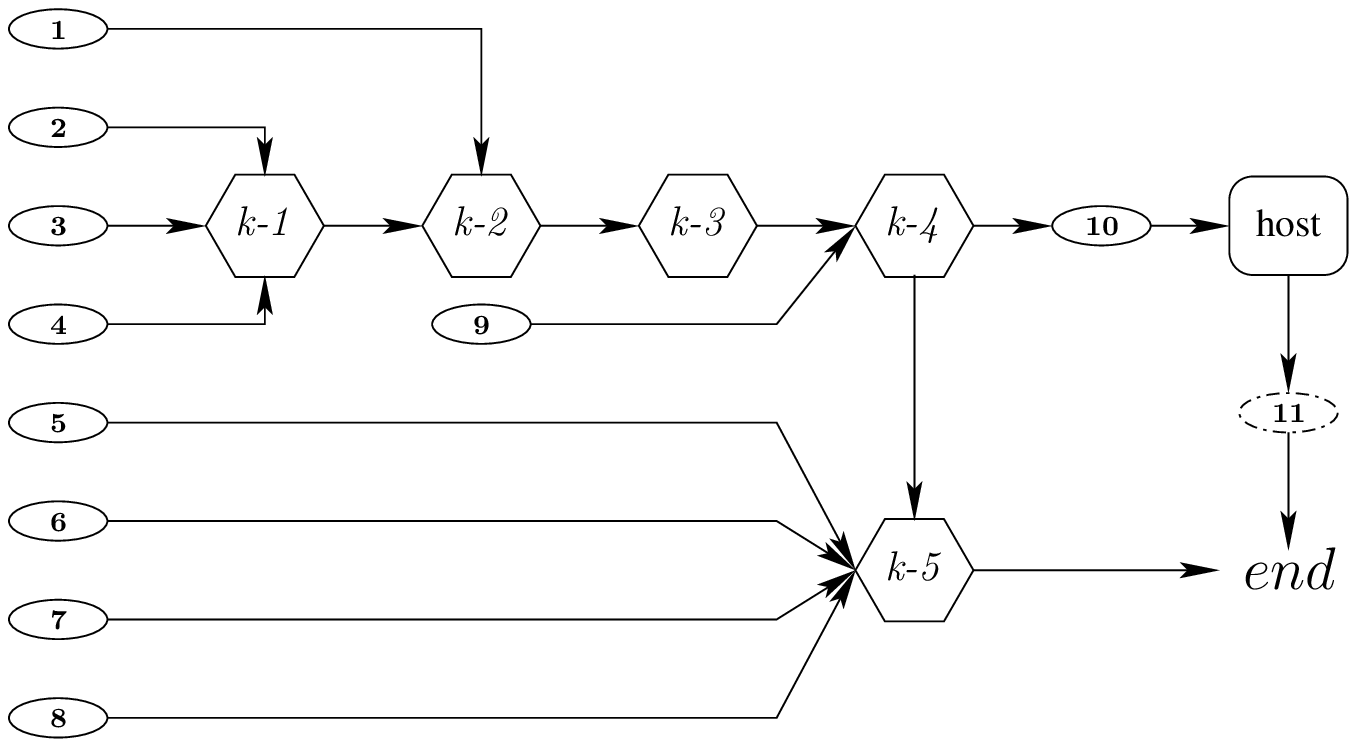}
{This exhibits the flow chart of jobs that will be fulfilled by
GPU and CPU at each simulation tick.
Arabic numbers represent various buffer reading and writing
tasks to be submitted to the OpenCL command queue.
\emph{k-1}, \emph{k-2}, ..., \emph{k-5} are the kernels
appearing in algorithm~\ref{a: opencl based gpu model}.
After task no.10, the host, i.e. CPU, will start to
process routine work.
Especially, if occupancy state of openings
is changed due to pedestrians' leaving and entering,
additional tasks, i.e. task no.11, will be submitted
correspondingly.}

\fig{width=1\textwidth}{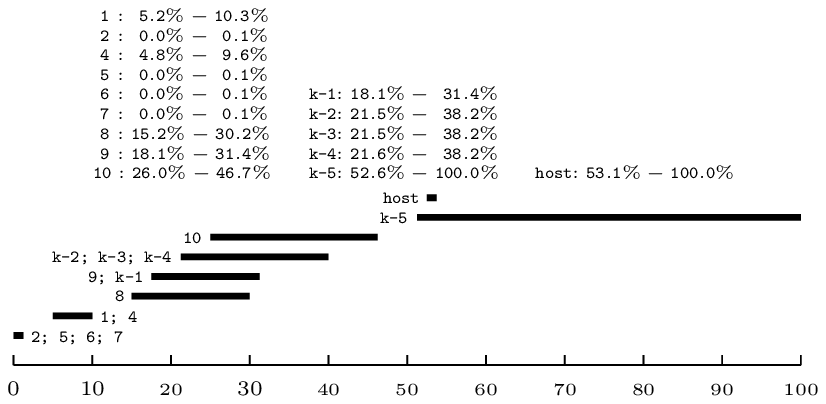}
{This exhibits each job's time frame of execution in percentage.
The ones for jobs no.3 and no.11 shown in figure~\ref{workflow.eps}
are missing due to usage of the periodic boundary condition.
This also causes the time frame used by the host to be very short,
which otherwise will largely overlap the one for job \textit{k-5}
in reality.  In addition, the shown ending time is the one
when the installed event callback procedure is called,
not exactly the one when the job is accomplished.
This is why multiple jobs appear to end at the same time.}

\item[\underline{\textit{Bank Conflict}}]
The way of emitting one machine instruction to
all work-items bundled in the same warp at a time
improves the degree of parallelism,
but also brings problems and difficulties deserving of attention.
One is the so called bank conflict.
In general, addresses of the local memory are grouped into banks.
For work-items in the same warp, access of the local memory
has to be serialized if the same bank is used.
Thus, to avoid bank conflicts, a feasible way is post-patch.
For example, in the current implementation,
internal \texttt{C struct\textrm{s}} used by work-items
will be intentionally patched to a size of a prime integer
by appending a corresponding number of bytes.
In this way, bank conflicts can be prevented at most.

\item[\underline{\textit{Divergence}}]
Another serious problem caused by the SIMD processing fashion
is divergence.  As all work-items in a warp will execute the same
machine instruction at a time, existence of branch statements
such as \texttt{if-else}, \texttt{switch} etc will result in
a portion of work-items to do idle work.
A general solution does not exist.
However some practices like using the trinary operator
\texttt{?:} if possible etc are better to be followed.
Thus the implementation is wholly re-structured
to reduce appearance of branch statements.
For instance, according to \cite{yu.et.al-2},
an array of eight strength will be sorted to determine
each pedestrian's next movement at each simulation tick.
The sorting was previously accomplished by using
the general quick-sort algorithm.
Now a tailored algorithm completely based on \texttt{?:}
is used instead in expense of generality and flexibility.

\end{description}

\tab{t: summary of computational techniques}
{This summarizes severity and implementation difficulty
of considered computational techniques.
A smaller value means more severe or harder.}{
\begin{tabular}{ccc}
 & Severity & Hardness\\\hline\noalign{\smallskip}
 \textit{Data Bandwidth} & \circled{1} & \circled{6}\\
 \textit{Competence} & \circled{2} & \circled{1}\\
 \textit{Divergence} & \circled{3} & \circled{2}\\
 \textit{Bank Conflict} & \circled{4} & \circled{4}\\
 \textit{Concurrency} & \circled{5} & \circled{5}\\
 \textit{Locality of Access} & \circled{6} & \circled{3}\\
\end{tabular}}

To map the social field model into
the OpenCL heterogeneous framework, factors
including those already discussed in the work of
\cite{yu.et.al-2} are considered so far.
Thus it is meaningful to have a summary
(table~\ref{t: summary of computational techniques}),
among which the issue of competence is more to say.
To solve competence without atomic functions,
two general methodologies exist.
The first one is to interweave operations
in a way so that no competition would occur,
for example \cite{komatitsch.et.al}.
Unfortunately, such an interweaving may not
exist for all of problems including
the one being discussed.
The second one follows the idea of
sacrifice of space in terms of time.
However it may cause memory to be exhausted quickly.
To solve, the idea of divide-and-conquer
can be resorted to set up an upper bound,
as what is exhibited in the paper.

\section{Numerical Experiments}
\label{s: numerical experiments}

\begin{figure}[ht]
 \begin{minipage}[t]{0.45\textwidth}
  \centering
  \includegraphics[width=\textwidth,scale=0.3]{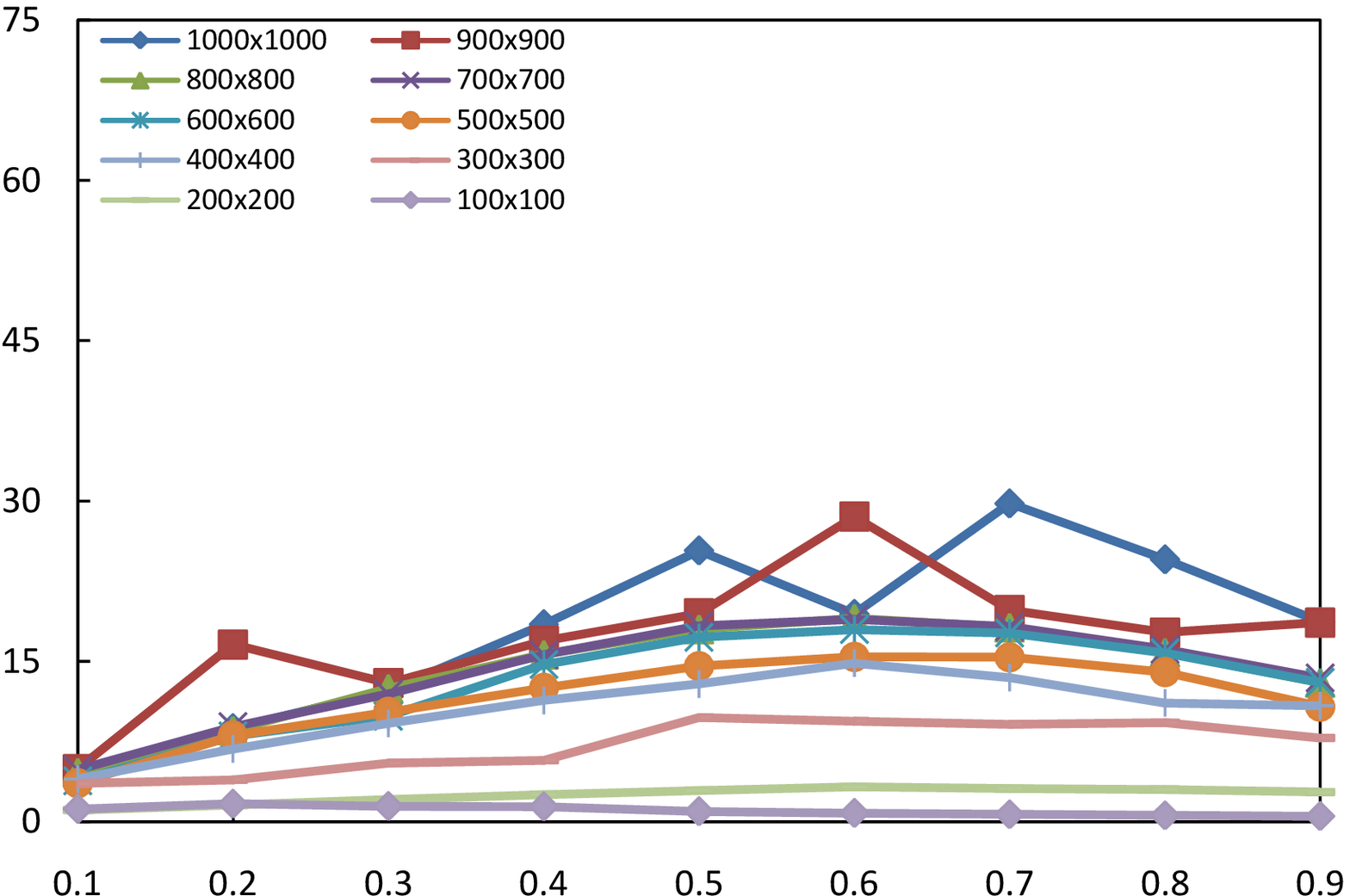}
 \end{minipage}
 \hfill
 \begin{minipage}[t]{0.45\textwidth}
 \centering
 \includegraphics[width=\textwidth,scale=0.3]{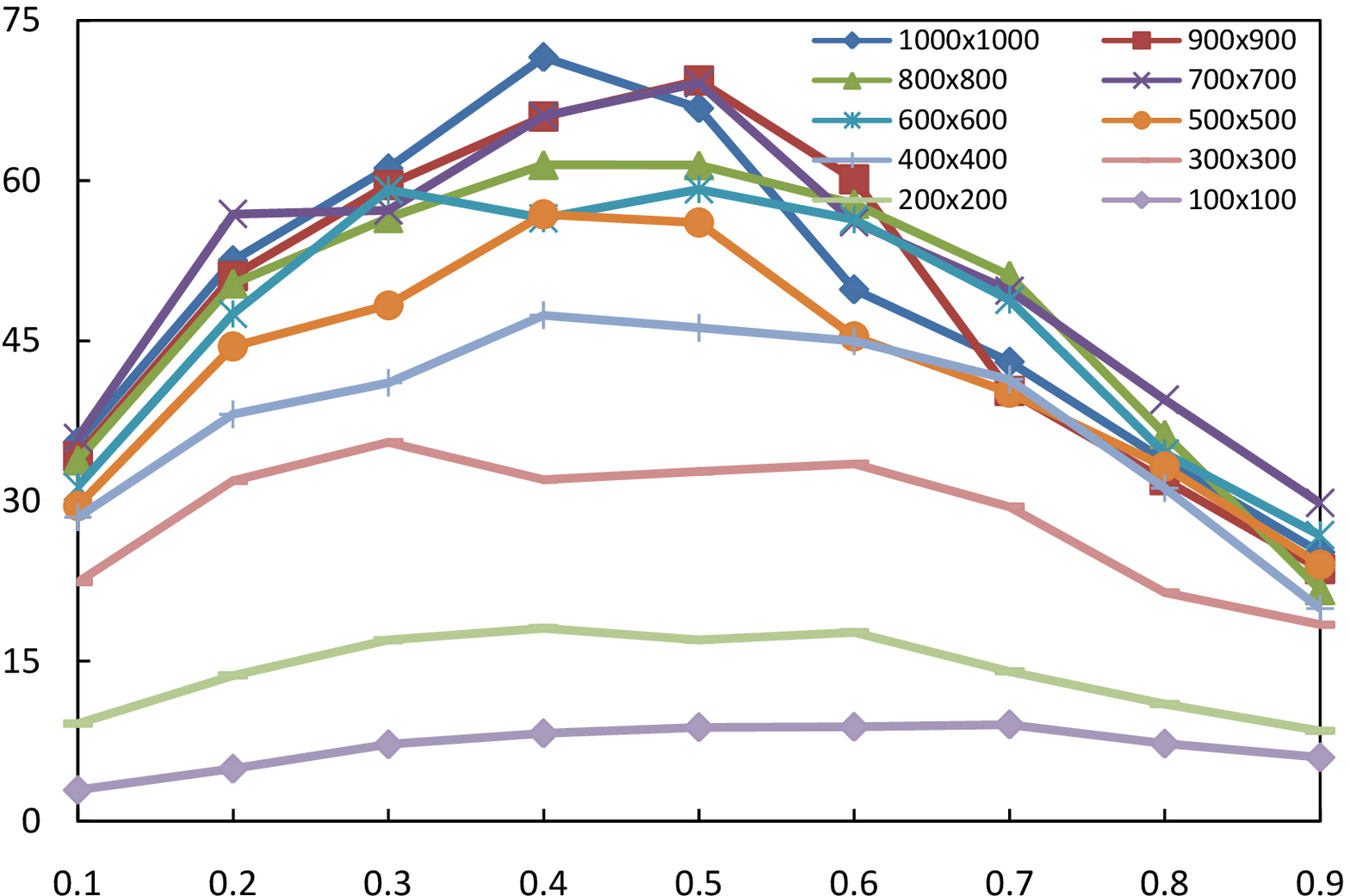}
 \end{minipage}
 \caption{\label{1-directional}%
  \small{This exhibits the performance ratios of running times of
  the CPU model to those of the GPU models for
  the uni-directional case.
  The left sub-plot shows the previous GPU model's and
  the right sub-plot shows the current GPU model's.}}
\end{figure}

\begin{figure}[ht]
 \begin{minipage}[t]{0.45\textwidth}
  \centering
  \includegraphics[width=\textwidth,scale=0.3]{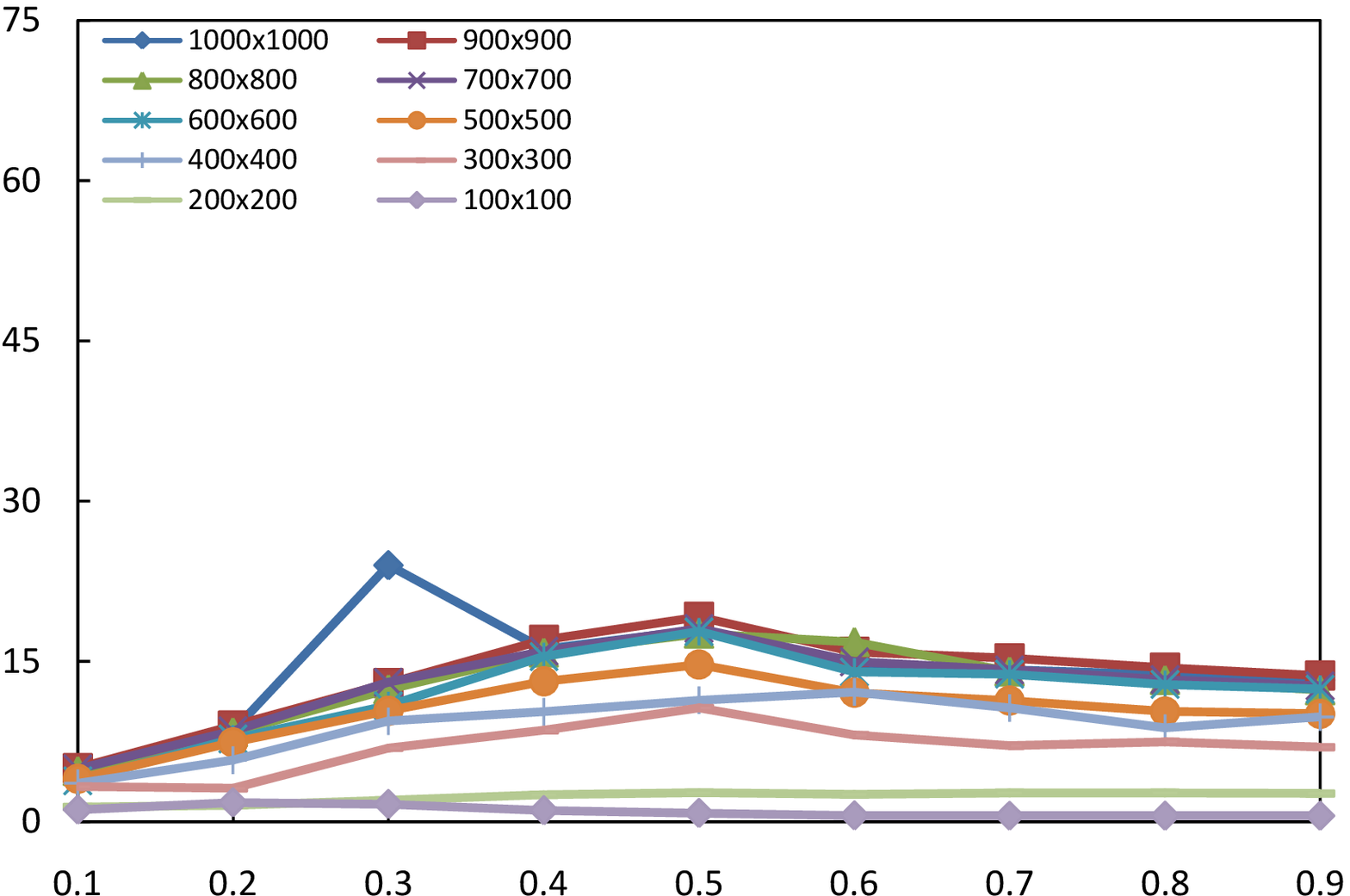}
 \end{minipage}
 \hfill
 \begin{minipage}[t]{0.45\textwidth}
 \centering
 \includegraphics[width=\textwidth,scale=0.3]{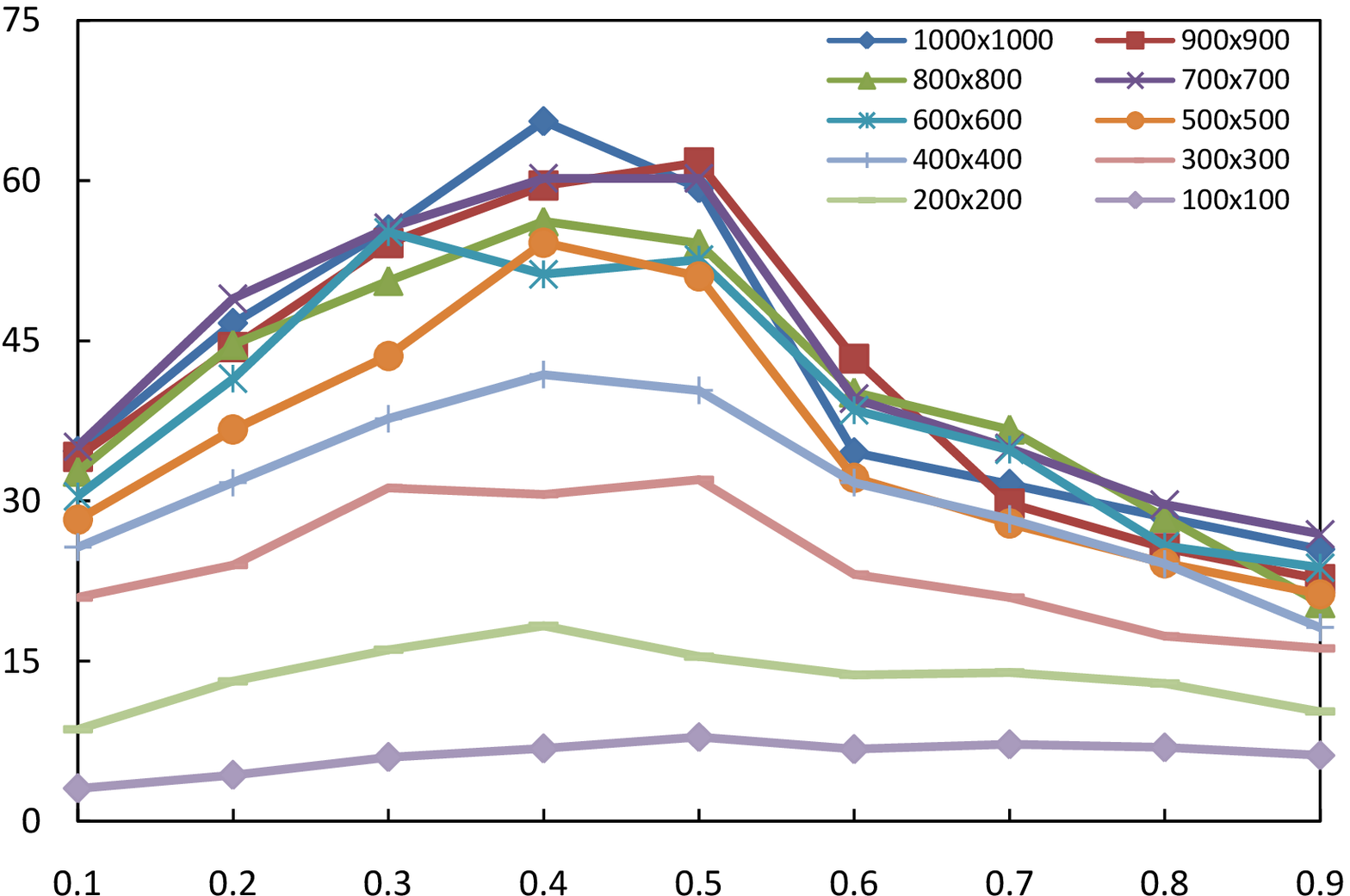}
 \end{minipage}
 \caption{\label{2-directional}%
  \small{Performance ratios (bi-directional)}}
\end{figure}

\begin{figure}[ht]
 \begin{minipage}[t]{0.45\textwidth}
  \centering
  \includegraphics[width=\textwidth,scale=0.3]{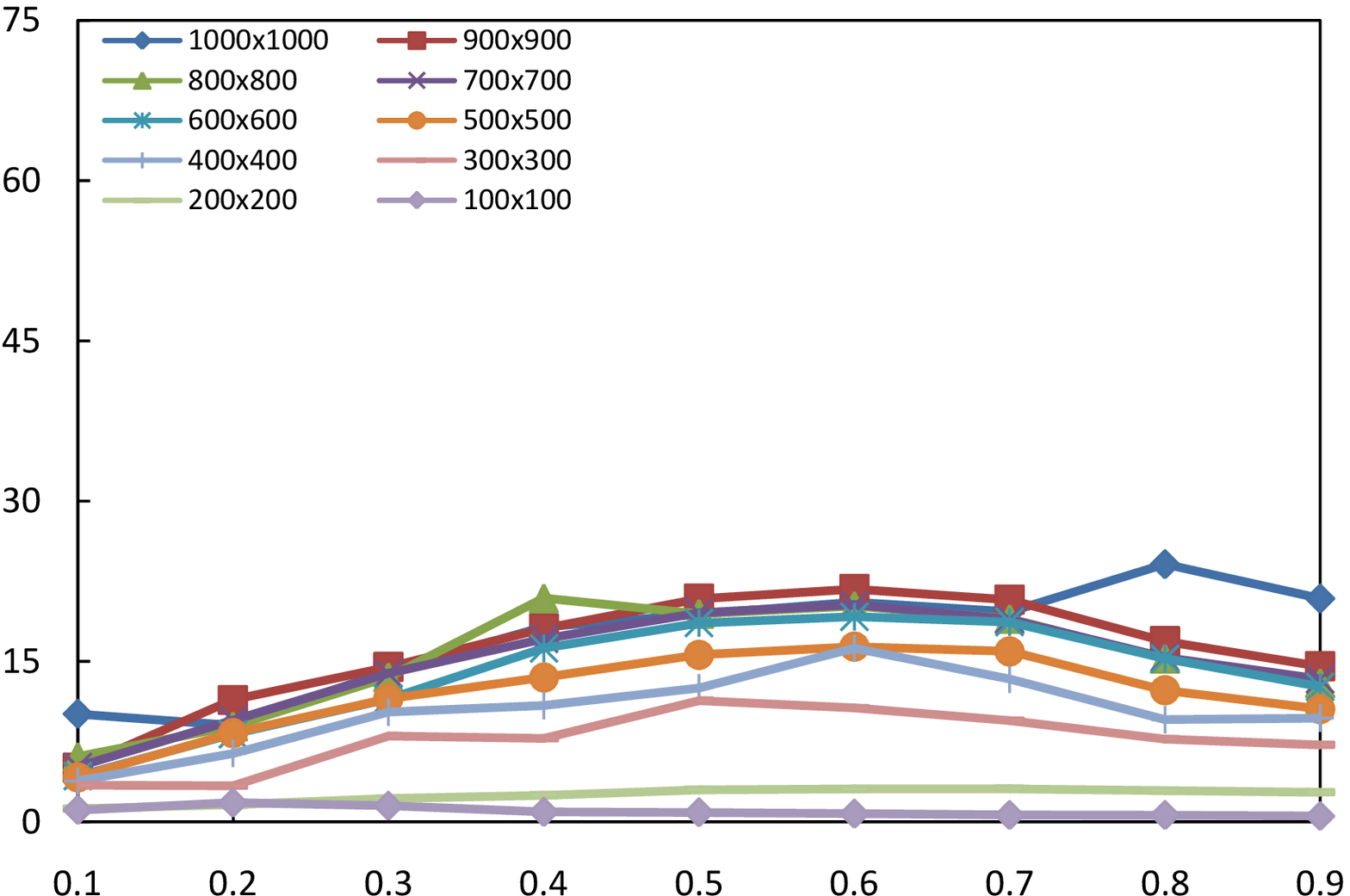}
 \end{minipage}
 \hfill
 \begin{minipage}[t]{0.45\textwidth}
 \centering
 \includegraphics[width=\textwidth,scale=0.3]{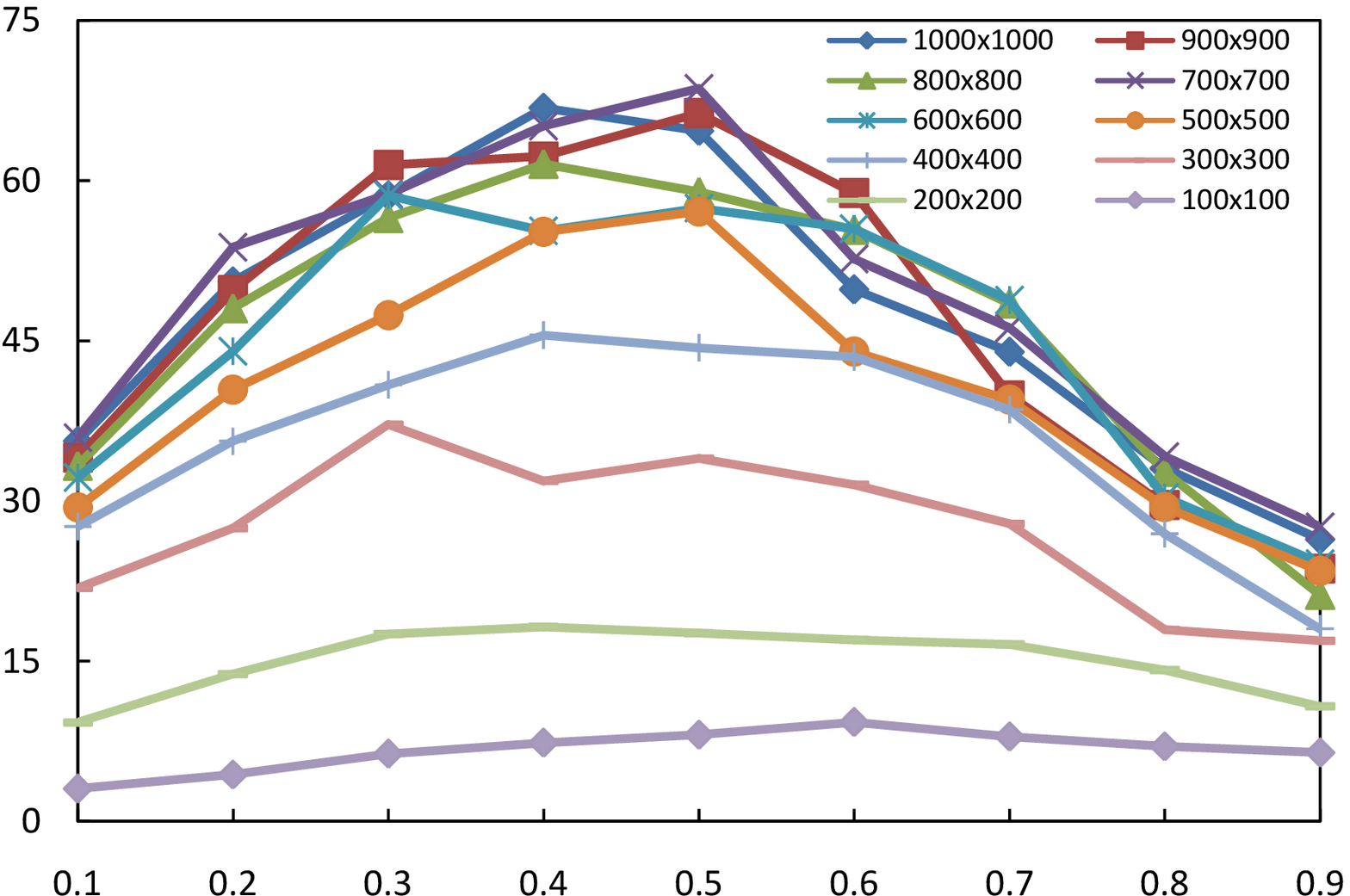}
 \end{minipage}
 \caption{\label{4-directional}%
  \small{Performance ratios (4-directional)}}
\end{figure}

\begin{figure}[ht]
 \begin{minipage}[t]{0.45\textwidth}
  \centering
  \includegraphics[width=\textwidth,scale=0.3]{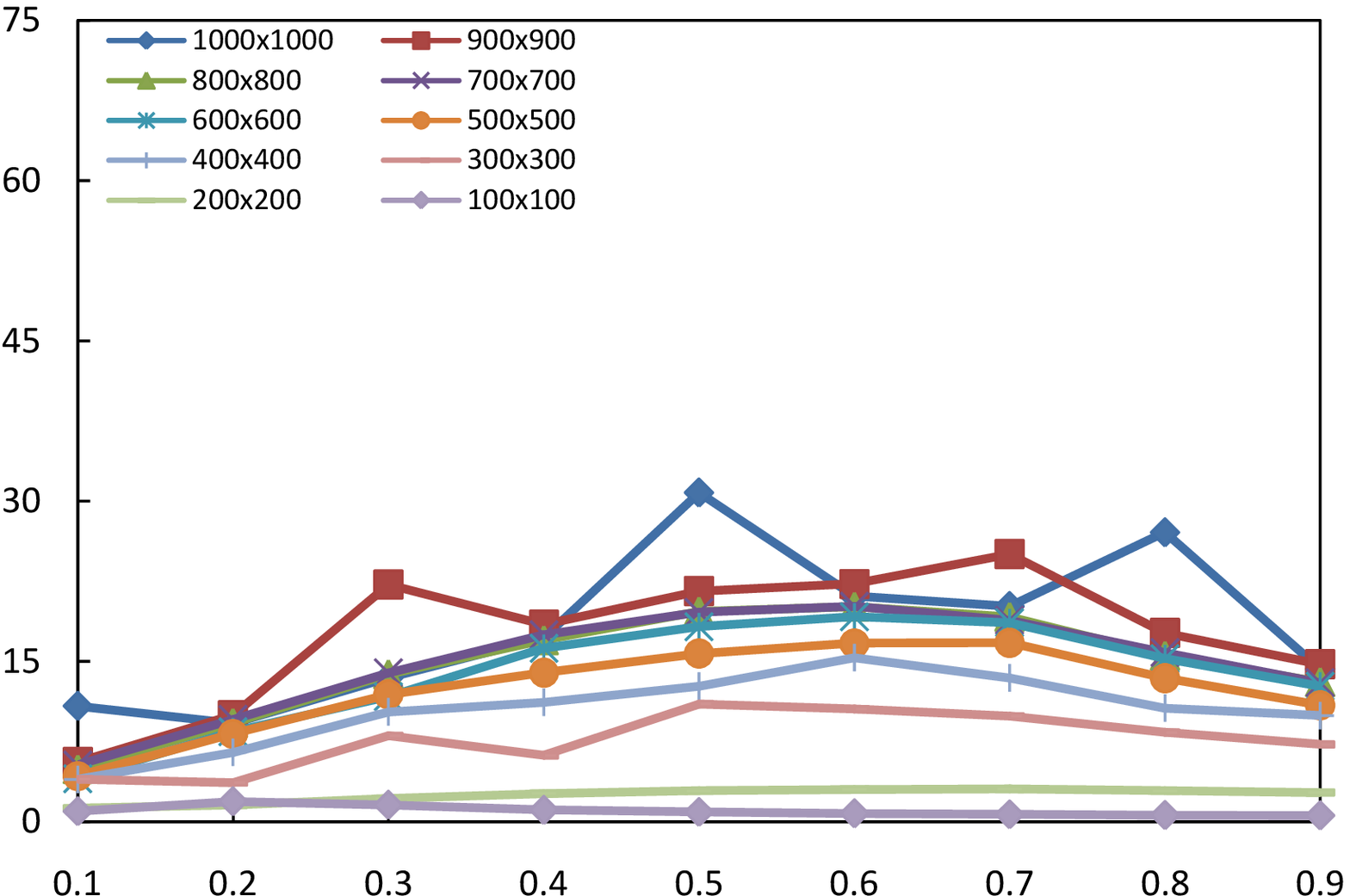}
 \end{minipage}
 \hfill
 \begin{minipage}[t]{0.45\textwidth}
 \centering
 \includegraphics[width=\textwidth,scale=0.3]{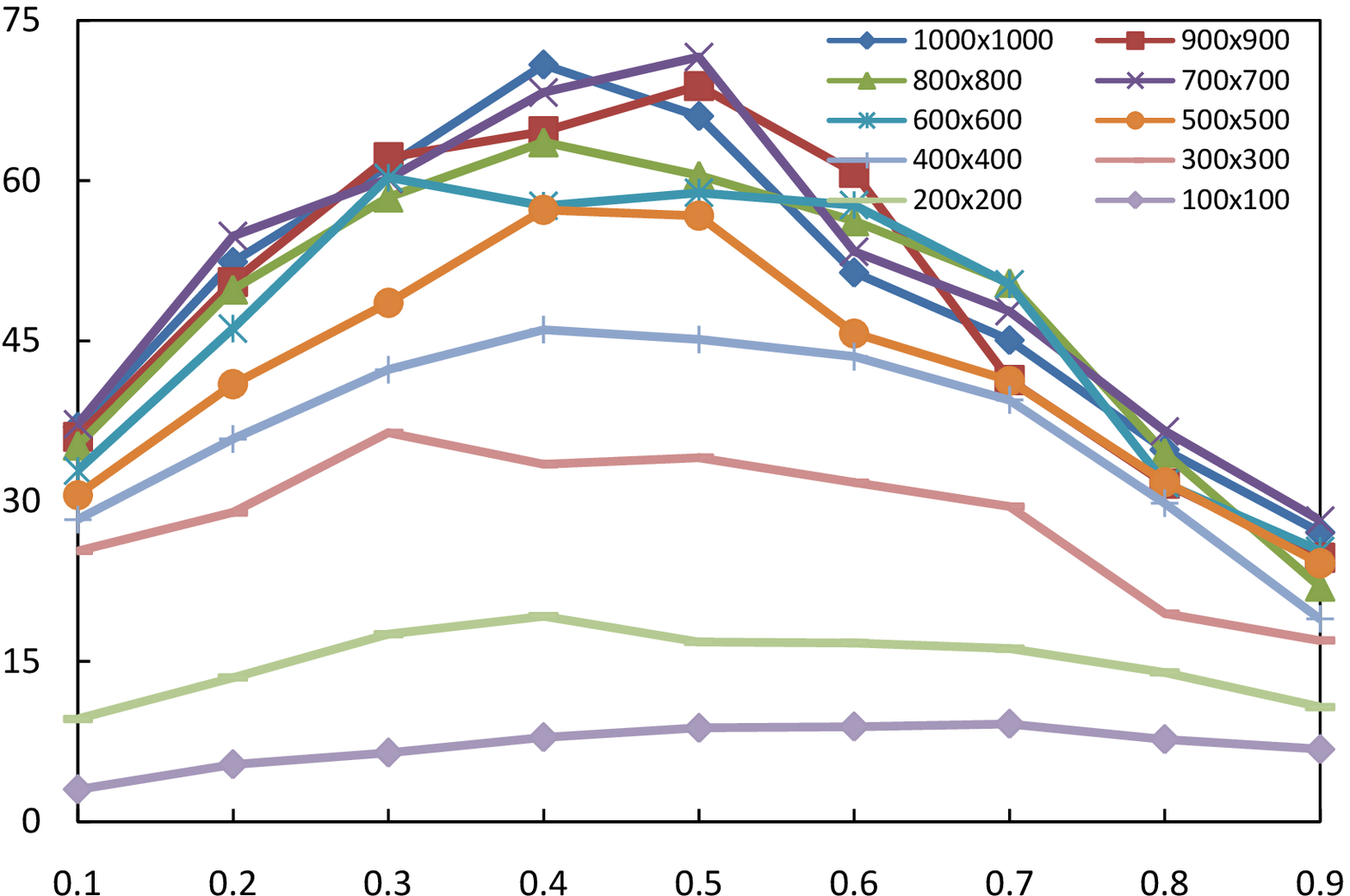}
 \end{minipage}
 \caption{\label{8-directional}%
  \small{Performance ratios (8-directional)}}
\end{figure}

In order to examine the current GPU model's numerical efficiency,
the scenarios used in the work of \cite{yu.et.al-2} are re-experimented.
Firstly, the discrete space runs from $100\times 100$,
$200\times 200$, $\cdots$, to $1000\times 1000$.
Secondly, for each discrete space, the macroscopic density
runs from $0.1$, $0.2$, $\cdots$, to $0.9$.
Thirdly, all fields' geometry is set to $7\times 7$.
Lastly, for each macroscopic density, the pedestrian flow
is assumed to be uni-directional, bi-directional,
4-directional, and 8-directional.
This gives total $10*9*4=360$ combinations.
Each combination lasts $1000$ simulation ticks and repeats
$10$ times to derive an average running time.
Especially, to keep the number of pedestrians constant,
the periodic boundary condition is used.
And pedestrians' geometry and walk period are all
assumed to $1\times 1$ and $1$.
Figures~\ref{1-directional}, \ref{2-directional},
\ref{4-directional} and \ref{8-directional} give
the comparison results.

As exhibited, the current GPU model's numerical efficiency
is even better.
With comparison to the CPU model, the previous GPU model's
performance improvement was between 15x and 25x most of the time.
And the highest value was 30.8x.
Now the current GPU model's performance improvement is
between 45x and 60x most of the time, comparing to the CPU model.
And the highest value 71.56x happens in the experimented scenario
(geometry $700\times 700$; density 0.5; 8-directional).
This can also be clearly seen in figure~\ref{former-revised.eps},
which plots all of the $360$ performance ratios of running times
of the previous GPU model to those of the current GPU model.
Furthermore it delineates that the average improvement ratio is 4.44x.

\fig{width=0.6\textwidth}{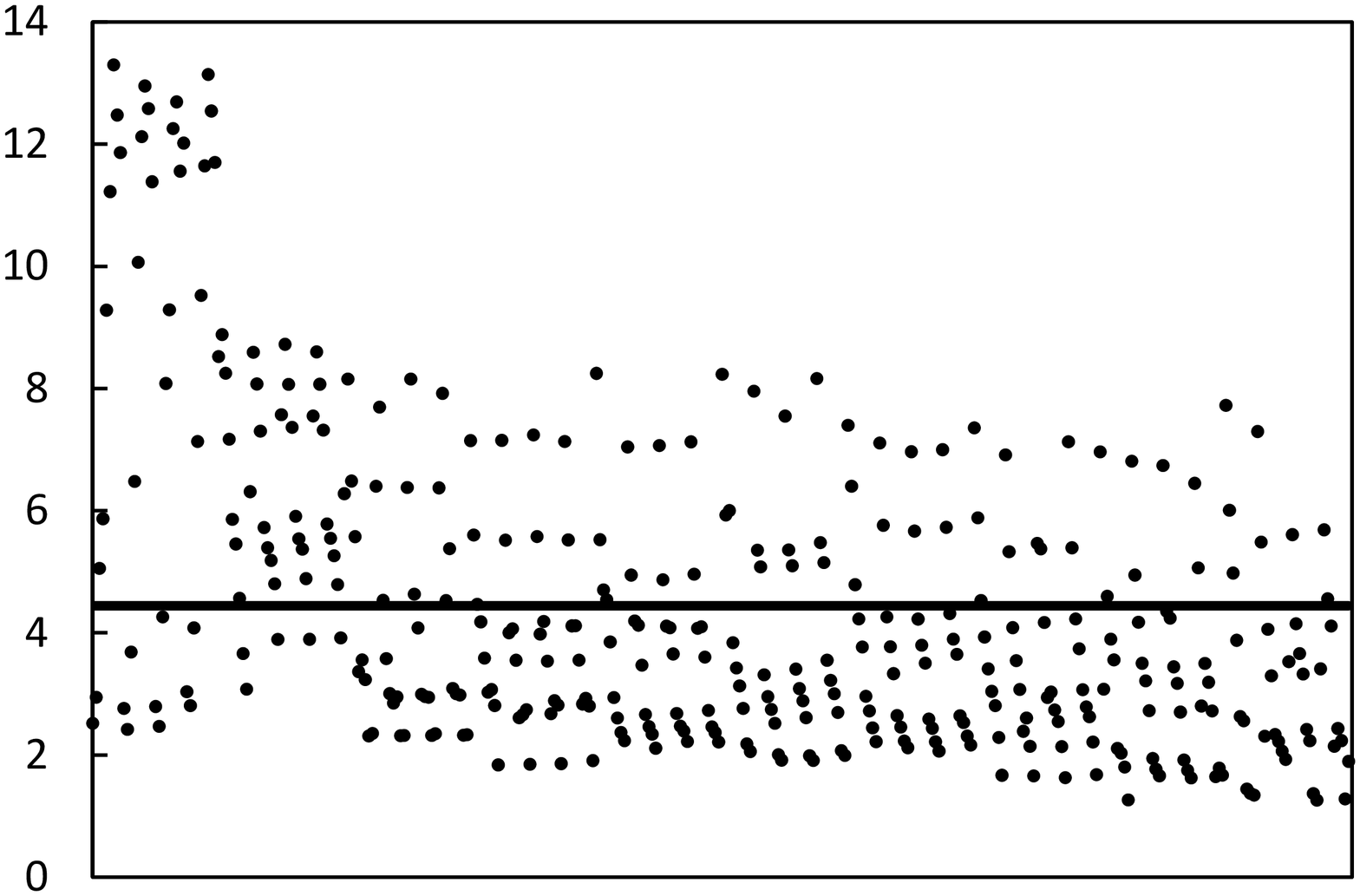}
{This exhibits the all $360$ performance ratios of
running times of the previous GPU model to
those of the current GPU model.
The maximal performance ratio is $13.3$
(geometry $100\times 100$; density 0.7; uni-directional).
The minimal performance ratio is $1.25$
(geometry $1000\times 1000$; density 0.9; 4-directional).
And the average value is $4.44$.}

\tab{t: multi-step sum}{Running Times}{
\begin{tabular}{c|c}
geometry & running time (seconds)\\\hline
$7\times 7$ & $62$\\
$21\times 21$ & $317$\\
$35 \times 35$ & $738$\\
$49 \times 49$ & $1424$\\
$63 \times 63$ & $2187$\\
$77 \times 77$ & $3238$\\
\end{tabular}}

\fig{width=0.6\textwidth}{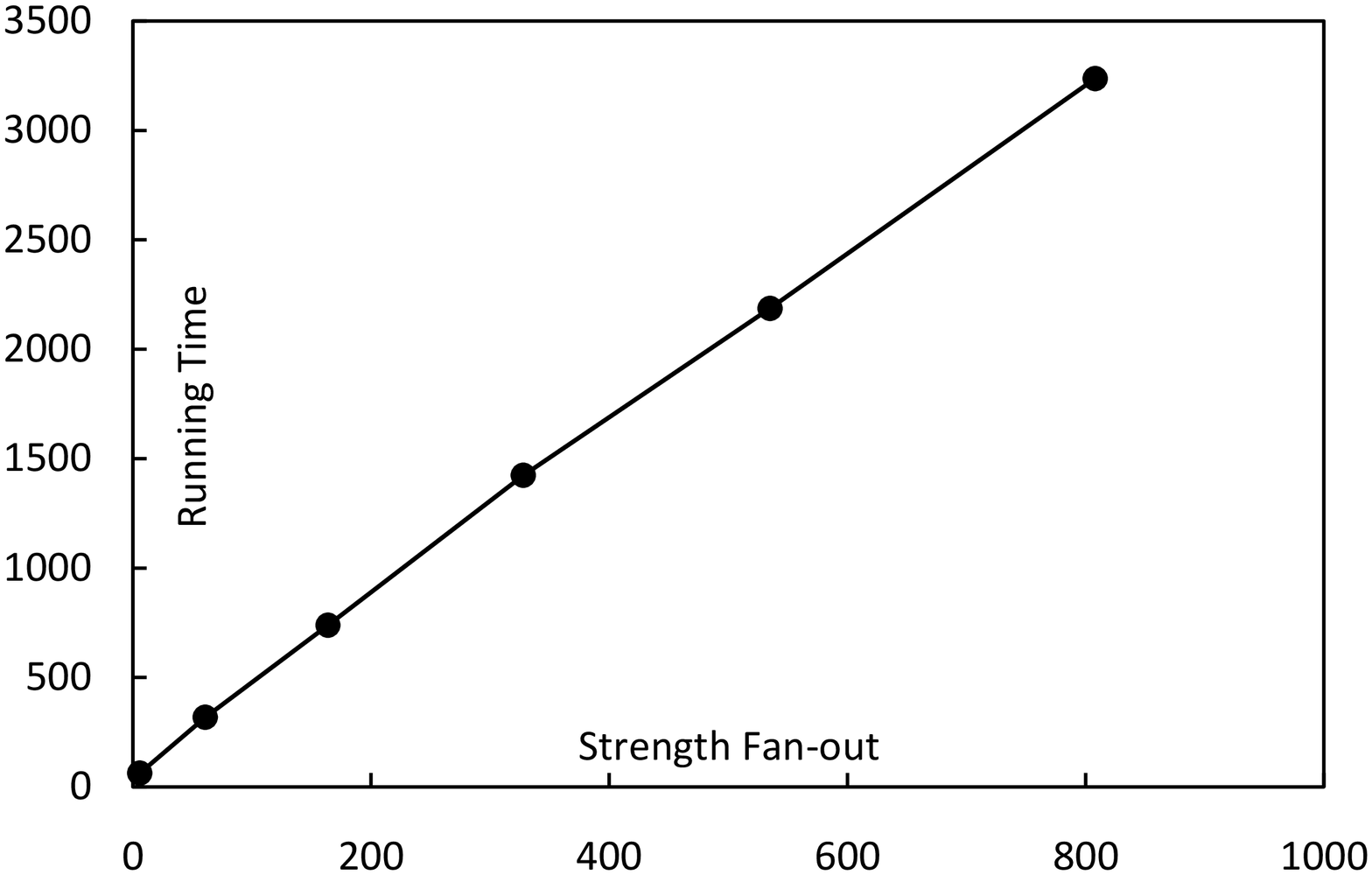}
{This exhibits the strict linearity between
the running time and strength fan-out.}

To see the efficiency of the algorithm introduced in
the section~\ref{s: solution illustration},
the scenario
(geometry $1000\times 1000$; density 0.5; 8-directional)
is chosen to be the baseline.
Then simulations are experimented by letting fields'
geometry running from $7\times 7$, $21\times 21$, ...,
to $77\times 77$.
The examined geometries match the ratios shown in
table~\ref{t: strength fan-out 2}.
Especially, since pedestrians' geometry is $1\times 1$
and a typical pedestrian's physical size is $0.3m \times 0.3m$,
this is equivalent to consideration of impact of
neighboring pedestrians, openings, and obstacles
within $0.3m$, $2.1m$, ..., till $23.1m$
in the pedestrian model.
The results are given in table~\ref{t: multi-step sum}.
In addition, according to algorithm~\ref{a: opencl based gpu model},
a strict linearity should exist between the running time
and strength fan-out, which is exhibited in
figure~\ref{linearity.eps}.

\tab{t: walk period}
{This exhibits running times in second for different walk periods.
Again the scenario (geometry $1000\times 1000$; density $0.5$;
8-directional) is chosen to be the baseline.
And simulations are accomplished by keeping the minimal period
being $1$ and letting the maximal period being $1$, $3$, ...,
and so on.
For convenience, it also indicates performance ratios
of the current GPU model with comparison to
the previous GPU model and the CPU model.}{
\begin{tabular}{c|cccccc}
Maximal Period & 1 & 3 & 5 & 7 & 9 & 11\\\hline
GPU model (current) & $62$ & $64$ & $64$ & $63$ & $62$ & $62$\\\hline
GPU model (previous) & $134$ & $203$ & $207$ & $207$ & $208$ & $208$\\
\textit{Ratio} & \textit{2.2} & \textit{3.3} & \textit{3.3} & \textit{3.3} & \textit{3.4} & \textit{3.4}\\\hline
CPU model & $4114$ & $2478$ & $2581$ & $2436$ & $2452$ & $2456$\\
\textit{Ratio} & \textit{66.4} & \textit{40.0} & \textit{41.6} & \textit{39.3} & \textit{39.5} & \textit{39.6}\\
\end{tabular}}

The concept of walk period was introduced for study of
varied walking velocities.
The method's most valuable advantage is that
no additional complexity will be added to
the underlying model logic.
In the work of \cite{yu.et.al-2}, impact of the mechanism upon
the numerical efficiency was experimented.
With improvements being introduced, the experiments are
rerun with the results being given in
table~\ref{t: walk period}.
Except the current GPU model wins again,
it is noticed that using a large maximal walk period
seems to have different impact on the three models.
For example, for the current GPU model,
the impact is negligible, but it will cause the previous GPU model
to use a long period to finish the simulation.

As aforementioned, in the discrete model,
the concept of walk period has an implicit relationship
with the space fineness so that pedestrians are allowed
to occupy more than one \su.  A by-product is that
the study of jostling is now feasible through
dynamically adjusting pedestrians' geometry.
Therefore an experiment considering both walk period
and dynamic geometry is conducted.
Samely the scenario (geometry $1000\times 1000$; density 0.5;
8-directional) is used as the baseline.
The walk periods run from 1, 3, ..., to 11
and pedestrian geometries run from
$1\times 1$, $3\times 3$, ..., to $11\times 11$.
Noteworthy, as the whole space is kept as $1000\times 1000$,
changing pedestrians' geometry
will impact the number of pedestrians and fields' geometry.
The results are given in table~\ref{t: combo}.
Firstly, it is seen that, for each pedestrian geometry considered,
the difference of running times for different walk periods
is insignificant.  This coincides with the conclusion made in
the previous experiment.
Secondly, it seems that fields' geometry has a more serious impact.
Increasing pedestrians' geometry
will both decrease the population and increase fields' geometry
in the almost same ratio.
But the running time becomes larger with comparing to
the basic case where pedestrians' geometry is $1\times 1$.
To some extent, this coincides with the fact that
\textit{k-5} is the most time consuming one among
the jobs submitted according to figure~\ref{percent.eps}.
Additionally, among the six pedestrian geometries experimented,
the one for $3\times 3$ requires the longest running time.

\tab{t: combo}{This exhibits running times for different
combinations of walk period and pedestrian geometry.
Meanwhile, for each experimented pedestrian geometry,
the corresponding number of pedestrians and fields' geometry
are also given.  For example, for pedestrian geometry $1\times 1$,
the number of pedestrians and fields' geometry are $500000$ and
$7\times 7$.}{
\begin{tabular}{c|cccccc}
 & $500000$ & $55555$ & $20000$ & $10204$ & $6172$ & $4132$\\
 & $7\times 7$ & $21\times 21$ & $35\times 35$ & $49\times 49$ & $63\times 63$ & $77\times 77$\\\cline{2-7}
 & $1\times 1$ & $3\times 3$ & $5\times 5$ & $7\times 7$ & $9\times 9$ & $11\times 11$\\\hline
1 & $63.5$ & $264.6$ & $213.5$ & $207.7$ & $197.6$ & $201.5$\\
3 & $63.9$ & $261.7$ & $202.0$ & $197.1$ & $186.9$ & $192.5$\\
5 & $64.0$ & $264.1$ & $204.6$ & $195.5$ & $184.4$ & $191.2$\\
7 & $63.8$ & $262.9$ & $201.3$ & $193.9$ & $182.0$ & $191.6$\\
9 & $63.7$ & $262.2$ & $200.1$ & $194.5$ & $184.2$ & $190.5$\\
11 & $63.9$ & $261.8$ & $200.8$ & $193.8$ & $182.1$ & $189.6$\\
\end{tabular}}

\section{Conclusion}
\label{s: conclusion}

In the paper, the previous OpenCL-based implementation of
the social field model is improved in two aspects.
In one aspect, the problem of memory depletion is solved
by the idea of divide-and-conquer.
The computational model is now ready to power analysis of
super-large scale crowd's complicated and finer walking behaviors.
In the other aspect, the OpenCL heterogeneous framework
is thoroughly studied and relevant computational techniques
are implemented, which brings the numerical efficiency
to an even higher level.

With regarding to the future work, first of all the authors
plan to develop useful transportation related functional modules,
such as XML-based complicated scenario abstraction,
macroscopic/mesoscopic route selection algorithms, and so on.
Secondly, the authors will set out to study how to use
the contemporary information technologies such as deep-learning etc
to automatically collect valuable data from raw video images.
At the moment, the lack of relevant accurate data is a big problem
for quantitative validation and calibration of pedestrian models.
The authors believe that the studies together will help scholars
to better understand the complex dynamics of evacuation processes.

\end{document}